\documentclass[10pt]{article}

\usepackage{a4wide}
\usepackage{amsmath,amssymb}
\usepackage{verbatim}

\title{\textbf{Essential self-adjointness of Wick squares in quasi-free Hadamard representations on curved
spacetimes.}}
\author{Ko Sanders\thanks{E-mail:
kosanders@uchicago.edu}\\
Enrico Fermi Institute\\
University of Chicago\\
5640 South Ellis Avenue\\
Chicago, IL 60637}
\date{5 April 2012}

\newtheorem{definition}{Definition}[section]
\newtheorem{theorem}[definition]{Theorem}
\newtheorem{proposition}[definition]{Proposition}
\newtheorem{corollary}[definition]{Corollary}
\newtheorem{lemma}[definition]{Lemma}
\newenvironment{proof*}{\smallskip\par\noindent\emph{Proof: }
 \ignorespaces}{\hfill$\Box$\smallskip\par\ignorespaces}
\newtheorem{remark}[definition]{Remark}

\newcommand{\map}[3]{\ensuremath{#1\!:\!#2\!\rightarrow\!#3}}
\newcommand{\C}{\ensuremath{\mathbb{C}}}
\newcommand{\R}{\ensuremath{\mathbb{R}}}

\newcommand{\N}{\ensuremath{\mathbb{N}}}

\newcommand{\Test}{\ensuremath{C^{\infty}}}

\newcommand{\alg}[1]{\ensuremath{\mathcal{#1}}}

\begin{document}
\maketitle

\begin{abstract}
We investigate whether a symmetric, second order Wick polynomial $T$ of a free scalar field, including derivatives, is essentially self-adjoint on the natural (Wightman) domain in a quasi-free (i.e.\ Fock space) Hadamard representation. Our results apply to arbitrary spacetime dimensions $d\ge 2$, but we do restrict our attention to the case where $T$ is smeared with a test-function from a particular class $\mathcal{S}$, namely the class of sums of squares of test-functions. (This class of smearing functions is smaller than the class of all non-negative test-functions -- a fact which follows from Hilbert's Theorem.) Combining techniques from microlocal and functional analysis we prove that $T$ is essentially self-adjoint if it is a Wick square (without derivatives). In the presence of derivatives we prove the weaker result that $T$ is essentially self-adjoint if its compression to the one-particle Hilbert space is essentially self-adjoint. For the latter result we use W\"ust's Theorem and an application of Konrady's trick in Fock space. In the presence of derivatives we also prove that one has some control over the spectral projections of $T$, by describing it as the strong graph limit of a sequence of essentially self-adjoint operators.
\end{abstract}

\section{Introduction}

In quantum physics an observable is given by a self-adjoint operator in a Hilbert space. For unbounded operators the condition of self-adjointness is somewhat technical and rather difficult to prove, but unfortunately this problem can hardly be circumvented in quantum field theory, where unbounded operators are ubiquitous. Moreover, in the case of quantum field theories on curved spacetimes the Hilbert space representation is not fixed a priori, but instead it depends on the choice of an algebraic state on an abstract $^*$-algebra of observables. When trying to prove self-adjointness results one therefore has to try and accommodate several choices: the choice of spacetime, the choice of state, which fixes the Hilbert space representation through the GNS-construction, and finally the choice of operator.

To our knowledge the previous literature has been rather restrictive in these three aspects. It concerns mostly the field itself or the Wick square and it relies either on the Wightman axioms in Minkowski spacetime or on the properties of a ground state of a free field \cite{Borchers+,Langerholc+1965,Baez1989}. The goal of this paper is to prove some results which aim for generalisations in all three aspects, making use of the microlocal spectrum condition \cite{Brunetti+3} and focussing on second order Wick polynomials with derivatives.\footnote{For remarks concerning the (generalised) free field itself in curved spacetime see \cite{Sanders}.} This includes such physically relevant operators as the components of the stress-energy-momentum tensor of a free scalar field. The essential self-adjointness of local energy densities (as a component of the stress tensor) would be an important step towards the development of a local and covariant version of the useful $H$-bounds \cite{Fredenhagen+1981} in Minkowski spacetime, which estimate the local singular behaviour of a quantum field in terms of the Hamiltonian operator $H$.

Let $T$ denote a second order Wick polynomial (with derivatives) on an arbitrary, globally hyperbolic spacetime of dimension $d\ge 2$, represented in a quasi-free Hadamard state. Then we will demonstrate the following results. Firstly we show in Lemma \ref{Lem_loc} that the problem at hand is local in the following sense: if $O\subset M$ is an open region that contains the supports of all the test-functions occurring in $T$, then $T$ restricts to the Hilbert space generated from $O$. If these restrictions are essentially self-adjoint for all relatively compact $O\subset M$, then $T$ itself is essentially self-adjoint too. Secondly, if $T$ is smeared with a function of class $\mathcal{S}$, the class of sums of squares of test-functions, then $T$ is essentially self-adjoint as soon as its compression to the one-particle Hilbert space is essentially self-adjoint. This is the content of Theorem \ref{Thm_SA}. Thirdly we show in Theorem \ref{Thm_SAGL} that if $T$ is smeared with a function of class $\mathcal{S}$, then $T$ has a self-adjoint extension which is the strong graph limit of a specific sequence of self-adjoint operators. This result provides a certain amount of control over the spectral projections of this self-adjoint extension. To actually prove essential self-adjointness we prove that the compression of $T$ to the one-particle Hilbert space is essentially self-adjoint when the operator $T$ is the Wick square (without derivatives) of a free scalar field, smeared with a real-valued test-function. Hence, when the test-function is in $\mathcal{S}$ the essential self-adjointness of the Wick square follows from Theorem \ref{Thm_SA}. This constitutes a substantial generalisation w.r.t.\ the class of spacetimes and representations considered in comparison with the literature.

We emphasise that our results do not make any use of the fact that the singularities in the two-point distribution of a free scalar field get weaker as $d$ decreases. An alternative strategy would be to consider small $d$ and try to exploit the weaker singularities. We comment on this strategy in our conclusions in Section \ref{Sec_Further}, especially in the extreme case $d=1$, in which case the free scalar field reduces to a harmonic oscillator and the Hadamard two-point distributions are smooth functions.

The remainder of this paper is organised as follows: after a review of some basic terminology and estimates for abstract Fock spaces in Section \ref{Sec_Fock}, we review in Section \ref{Sec_GenSA} the self-adjointness results that we will need in an abstract setting. In particular we prove a useful application of Konrady's Trick to Fock space and a Monotone Graph Limit Theorem. In Section \ref{Sec_QFHad} we set the stage for our physical applications by reviewing the GNS-representations of a (quasi-free) Hadamard state of a (generalised) free field and defining its Wick polynomials. Section \ref{Sec_SA} contains the main results described above and their proofs and includes a remark on the class $\mathcal{S}$ of smearing functions. We conclude our paper with some discussion in Section \ref{Sec_Further}.

\section{Review of Fock space}\label{Sec_Fock}

Because quasi-free states lead to representations with a Fock space structure we will first recall some information on Fock spaces that will be needed later on. Most of the following constructions are entirely standard and follow Section 5.2 of \cite{Bratteli+}.

Let $\mathcal{K}$ be a complex Hilbert space with inner product $\langle,\rangle$ (complex linear in the second entry). The (full, unsymmetrised) Fock space is defined as the direct sum of the Hilbert tensor products
\[
\mathcal{H}:=\bigoplus_{n=0}^{\infty}\mathcal{H}^{(n)},\quad
\mathcal{H}^{(n)}:=\mathcal{K}^{\otimes n},
\]
where $\mathcal{K}^{\otimes 0}:=\C$. We call $\mathcal{H}^{(n)}$ the $n$-particle subspace and we denote by $\mathcal{F}$ the subspace of finite particle vectors, i.e.\ the elements $\oplus_{n=0}^{\infty}\psi_n$ with only finitely many non-zero summands. We let $P_n$ denote the orthogonal projection operators onto the summands $\mathcal{H}^{(n)}$.

For each $u\in\mathcal{K}$ one defines a linear creation operator $\map{a^*(u)}{\mathcal{F}}{\mathcal{F}}$ and annihilation operator $\map{a(u)}{\mathcal{F}}{\mathcal{F}}$ by
\begin{eqnarray}\label{def_aops}
a^*(u)(v_1\otimes\ldots\otimes v_n)&:=&\sqrt{n+1}u\otimes v_1\otimes\ldots\otimes v_n\nonumber\\
a(u)(v_1\otimes\ldots\otimes v_n)&:=&\sqrt{n}\langle u,v_1\rangle v_2\otimes\ldots\otimes v_n\nonumber
\end{eqnarray}
on $\mathcal{H}^{(n)}$ and $a(u)|_{\mathcal{H}^{(0)}}:=0$. Both are densely defined linear operators in $\mathcal{H}$. Notice that
\[
\|a^*(u)P_n\|=\sqrt{n+1}\|u\|_{\mathcal{K}},\quad \|a(u)P_n\|=\sqrt{n}\|u\|_{\mathcal{K}},
\]
where the first follows from the definition and the second from the fact that $a(u)P_{n+1}$ is the adjoint of the bounded operator $a^*(u)P_n$.

The symmetric (or bosonic) Fock space is defined similarly as
\[
\mathcal{H}_+:=\bigoplus_{n=0}^{\infty}\mathcal{H}_+^{(n)},\quad
\mathcal{H}_+^{(n)}:=\mathcal{K}^{\otimes_s n},
\]
where $\otimes_s$ denotes the symmetrised tensor product. It is often convenient to view $\mathcal{H}_+^{(n)}\subset\mathcal{H}^{(n)}$ as a subspace, with a canonical projection operator $P_{+,n}$ defined as the bounded linear extension of
\[
P_{+,n}(v_1\otimes\ldots\otimes v_n):=\frac{1}{n!}\sum_{\pi\in S_n}v_{\pi(1)}\otimes\ldots\otimes v_{\pi(n)},
\]
where $S_n$ is the group of permutations of the set $\left\{1,\ldots,n\right\}$. Similarly, $\mathcal{H}_+$ can be viewed as a subspace of $\mathcal{H}$ with the projection operator $\map{P_+}{\mathcal{H}}{\mathcal{H}_+}$ defined as $P_+:=\oplus_{n=0}^{\infty}P_{+,n}$.

Annihilation and creation operators on the symmetric Fock space are defined on the dense domain $\mathcal{F}_+:=P_+\mathcal{F}$ in $\mathcal{H}_+$ by identifying
\[
a_+(u):=a(u)P_+,\quad a^*_+(u):=P_+a^*(u)P_+,
\]
where we notice that $a(u)$ preserves the symmetry. The symmetrisation also enforces the commutation relations $[a_+(u),a_+^*(v)]=\langle u,v\rangle I$, and $[a_+(u),a_+(v)]=0=[a^*_+(u),a^*_+(v)]$ for any $u,v\in\mathcal{K}$.

As $P_{+,n}=P_+P_n=P_nP_+$ we have:
\begin{eqnarray}
\|a_+(u)P_{+,n}\|&\le&\|a(u)P_n\|=\sqrt{n}\|u\|_{\mathcal{K}},\nonumber\\
\|a^*_+(u)P_{+,n}\|&\le&\|a^*(u)P_n\|=\sqrt{n+1}\|u\|_{\mathcal{K}}.\nonumber
\end{eqnarray}
The following more detailed estimate\footnote{This is a straightforward generalisation of part of Theorem X.44 in \cite{Reed+}, who only consider $\mathcal{K}=L_2(\R^3)$.} will be of some use in subsequent sections. In it we use
$\overline{\mathcal{K}}$ to denote the conjugate Hilbert space of a Hilbert space $\mathcal{K}$, so there is a complex anti-linear bijection $\map{j}{\overline{\mathcal{K}}}{\mathcal{K}}$ such that
$\langle u,v\rangle=\langle j(v),j(u)\rangle$. Note that $\overline{\mathcal{K}}_1\otimes\overline{\mathcal{K}}_2=\overline{\mathcal{K}_1\otimes\mathcal{K}_2}$.
\begin{proposition}\label{Prop_Nest}
For $n\ge m$ the multilinear map
\[
\tilde{A}^{(l,m;n)}:(u_1,\ldots,u_l,v_1,\ldots,v_m)\mapsto a^*_+(u_1)\cdots a^*_+(u_l)a_+(j(v_1))\cdots a_+(j(v_m))P_{+,n}
\]
from $\mathcal{K}^{\times l}\times\overline{\mathcal{K}}^{\times m}$ to the space $\mathcal{B}(\mathcal{H}_+)$ of bounded linear operators on the symmetric Fock space gives rise (by definition of the tensor product) to a unique linear map $\map{A^{(l,m;n)}}{\mathcal{K}^{\otimes l}\otimes\overline{\mathcal{K}}^{\otimes m}}{\mathcal{B}(\mathcal{H}_+)}$ which is bounded, with $\|A^{(l,m;n)}\|\le\frac{\sqrt{n!(n-m+l)!}}{(n-m)!}$.
\end{proposition}
The operators in the range of $A^{(l,m;n)}$ are Hilbert-Schmidt operators from $\mathcal{H}_+^{(n)}$ to $\mathcal{H}_+^{(n-m+l)}$.
\begin{proof*}
First we notice that without symmetrisation we have for all $N,l$ and all $u^{(i)}_j\in\mathcal{K}$:
\begin{eqnarray}
\left\|\sum_{i=1}^Na^*(u^{(i)}_1)\cdots a^*(u^{(i)}_l)P_{+,n}\right\|^2&=&
\frac{(n+l)!}{n!}\left\|\sum_{i=1}^Nu^{(i)}_1\otimes\cdots\otimes u^{(i)}_l
\right\|^2_{\mathcal{K}^{\otimes l}},\nonumber
\end{eqnarray}
by direct computation. For any finite sum $F=\sum_i u^{(i)}_1\cdots u^{(i)}_l$ we use
$(a_+^*)^{\otimes l}(F)=P_+(a^*)^{\otimes l}(F)P_+$ and $\|P_+\|\le 1$ to find
\[
\|(a_+^*)^{\otimes l}(F)P_{+,n}\|\le\sqrt{\frac{(n+l)!}{n!}}\|F\|_{\mathcal{K}^{\otimes l}}.
\]
By continuous extension this must hold for all $F\in\mathcal{K}^{\otimes l}$. If
$G:=\sum_i v^{(i)}_1\cdots v^{(i)}_m\in\overline{\mathcal{K}}^{\otimes m}$ and $G':=\sum_i v^{(i)}_m\cdots v^{(i)}_1$ (with reversed order), then $\|G\|_{\overline{\mathcal{K}}^{\otimes m}}=\|G'\|_{\overline{\mathcal{K}}^{\otimes m}}
=\|j^{\otimes m}G'\|_{\mathcal{K}^{\otimes m}}$. Taking the adjoint $(a_+^*(u)P_n)^*=a_+(u)P_{n+1}$ we therefore have
\[
\|a_+^{\otimes m}(G)P_{+,n+m}\|=\|(a_+^*)^{\otimes m}(j^{\otimes m}G')P_{+,n}\|\le
\sqrt{\frac{(n+m)!}{n!}}\|G\|_{\overline{\mathcal{K}}^{\otimes m}}.
\]
We have now proved the result for the cases $m=0$ or $l=0$. For the general case we let
$F\in\mathcal{K}^{\otimes l}\otimes\overline{\mathcal{K}}^{\otimes m}$ and write it in terms of a Schmidt decomposition,\footnote{Such a Schmidt decomposition exists: let $F\in\mathcal{H}_1\otimes\mathcal{H}_2$, where the $\mathcal{H}_i$ are Hilbert spaces. Let $\left\{f_i\right\}_{i\in\mathcal{I}}$ be an orthonormal basis for $\mathcal{H}_1$ ($\mathcal{I}$ some index set) and define $g_i\in\mathcal{H}_2$ by
$\langle F,f_i\otimes\psi\rangle=\langle g_i,\psi\rangle$ for all $\psi\in\mathcal{H}_2$. Notice that $g_i$ vanishes for all but a countable number of indices $i$, by construction of the tensor product. Indexing the subset where $g_i\not=0$ by $i\in\N$ and setting $F_n:=\sum_{i=1}^nf_i\otimes g_i$ one proves $\langle F-F_n,F_n\rangle=0$, $\mathrm{w}-\lim_{n\rightarrow\infty}F_n=F$ and hence $F=\lim_{n\rightarrow\infty}F_n$.} $F=\sum_{i=1}^{\infty}f_i\otimes g_i$, where the $f_i\in\mathcal{K}^{\otimes l}$ are orthonormal and
$g_i\in\overline{\mathcal{K}}^{\otimes m}$. We compute for any $\psi\in\mathcal{H}_+^{(n)}$:
\begin{eqnarray}
\left\|\sum_{i=1}^{\infty}(a_+^*)^{\otimes l}(f_i)a_+^{\otimes m}(g_i)\psi\right\|^2&=&
\left\|P_+\sum_{i=1}^{\infty}(a^*)^{\otimes l}(f_i)a^{\otimes m}(g_i)\psi\right\|^2\le
\left\|\sum_{i=1}^{\infty}(a^*)^{\otimes l}(f_i)a^{\otimes m}(g_i)\psi\right\|^2\nonumber\\
&=&\frac{(n-m+l)!}{(n-m)!}\sum_{i,j=1}^{\infty}\langle f_i,f_j\rangle_{\mathcal{K}^{\otimes l}}\cdot
\langle a^{\otimes m}(g_i)\psi,a^{\otimes m}(g_j)\psi\rangle\nonumber\\
&=&\frac{(n-m+l)!}{(n-m)!}\sum_{i=1}^{\infty}\|f_i\|_{\mathcal{K}^{\otimes l}}^2\cdot
\|a^{\otimes m}(g_i)\psi\|^2\nonumber\\
&\le&\frac{n!(n-m+l)!}{(n-m)!^2}\sum_{i=1}^{\infty}\|f_i\|^2_{\mathcal{K}^{\otimes l}}\cdot
\|g_i\|^2_{\overline{\mathcal{K}}^{\otimes m}}\cdot\|\psi\|^2\nonumber\\
&=&\frac{n!(n-m+l)!}{(n-m)!^2}\|F\|^2_{\mathcal{K}^{\otimes l}\otimes\overline{\mathcal{K}}^{\otimes m}}
\cdot\|\psi\|^2\nonumber
\end{eqnarray}
which proves the desired estimate for $A^{(l,m;n)}$.
\end{proof*}

We call a polynomial in creation and annihilation operators normally ordered when all annihilation operators occur to the right of all creation operators. We remark that for non-normally ordered expressions the factor in Proposition \ref{Prop_Nest} may change, but is at most $\frac{(n+l)!}{\sqrt{n!(n-m+l)!}}$.

\section{General self-adjointness results}\label{Sec_GenSA}

In this section we collect some mathematical results that allow us to conclude that an operator is (essentially)
self-adjoint. In order to formulate these results we first need to introduce some further concepts.

Let $\mathcal{H}$ be a complex Hilbert space and $\map{X}{\mathcal{D}}{\mathcal{H}}$ a linear operator with a dense
domain $\mathcal{D}\subset\mathcal{H}$.
\begin{definition}\label{Def_AVectors}
A vector $\psi\in\mathcal{H}$ is called an \emph{analytic vector} for $X$ if and only if there exists a constant $c>0$
such that for all $n\in\N$ we have $X^n\psi\in\mathcal{D}$ and $\|X^n\psi\|\le c^{n+1}n!$.
\end{definition}
The set of analytic vectors for $X$ is a vector space and if $X$ is bounded then all vectors are analytic.

Now we have (see e.g.\ \cite{Reed+} Theorem X.39):
\begin{theorem}[Nelson's Theorem]\label{Thm_Nelson}
A symmetric operator $X$ on a Hilbert space, whose domain contains a dense set of analytic vectors, is essentially
self-adjoint.
\end{theorem}
A direct application of Nelson's Theorem to the Fock space setting is:
\begin{proposition}\label{Prop_NelsonFock}
Let $T=\sum_{n=0}^{\infty}\sum_{0\le l\le p\le 2}A^{(l,p-l;n)}(F^{(l,p-l)})$ be an operator on a symmetric Fock space $\mathcal{H}_+$, which is a normally ordered polynomial of order $\le 2$ smeared with elements
$F^{(l,m)}\in\mathcal{K}^{\otimes l}\otimes\overline{\mathcal{K}}^{\otimes m}$ and defined as a strong limit on $\mathcal{F}_+$. Then every vector in $\mathcal{F}_+$ is an analytic vector for $T$ and if $T$ is symmetric then it is essentially self-adjoint on $\mathcal{F}_+$.
\end{proposition}
\begin{proof*}
First notice that $T$ is well-defined by Proposition \ref{Prop_Nest}. Now define the subspaces $\mathcal{L}^{(n)}:=\bigoplus_{m=0}^n\mathcal{H}^{(m)}_+$ in $\mathcal{H}_+$ and notice that $T\mathcal{L}^{(n)}\subset\mathcal{L}^{(n+2)}$ for all $n\ge 0$. Again by Proposition \ref{Prop_Nest} we can estimate $\|T|_{\mathcal{L}^{(n)}}\|\le c(n+2)$ for some positive constant $c$ which is independent of $n$. Repeated application of this estimate yields
\[
\|T^m|_{\mathcal{L}^{(n)}}\|\le c^m\Pi_{j=0}^{m-1}(n+2j+2)\le c^m(n+2)^m m!,
\]
because $n+2j+2\le (n+2)(j+1)$. This proves that all vectors in the domain are analytic and essential self-adjointness follows from Nelson's Theorem.
\end{proof*}

A general strategy to conclude (essential) self-adjointness is to consider perturbations or suitable limits of operators which are already known to be (essentially) self-adjoint. A useful result for perturbations is the following
(cf.\ \cite{Reed+} Theorem X.14)
\begin{theorem}[W\"ust's Theorem]\label{Thm_Wuest}
Let $X$ be an essentially self-adjoint operator on a Hilbert space $\mathcal{H}$ with dense domain $\mathcal{D}$ and let $Y$ be a symmetric operator with domain $\mathcal{D}$ satisfying
\[
\|Y\psi\|\le \|X\psi\|+b\|\psi\|
\]
for some $b\ge 0$ and all $\psi\in\mathcal{D}$. Then $X+Y$ is essentially self-adjoint on $\mathcal{D}$.
\end{theorem}
Actually, the formulation of this theorem in \cite{Reed+} requires $X$ to be self-adjoint (i.e.\ $X$ must also be closed). However, given the estimate on $\mathcal{D}$, both operators $X$ and $Y$ can be uniquely extended to the domain of $\overline{X}$ and the estimate remains valid.

Unfortunately it is sometimes hard to prove that the perturbation $Y$ is small with respect to $X$, as W\"ust's Theorem requires. For this purpose we will apply an argument, known as Konrady's Trick (\cite{Reed+} Section X.2), which applies to the Fock space setting as follows:
\begin{theorem}\label{Thm_Konrady}
Let $N$ denote the number operator, defined on $\mathcal{F}_+$ by $N|_{\mathcal{H}_+^{(n)}}:=nI$. Let $X_n$ be a positive, essentially self-adjoint operator on $\mathcal{H}^{(n)}_+$ with dense domain $\mathcal{D}_n$ and let $X:=\bigoplus_{n=0}^{\infty}X_n$ be the essentially self-adjoint operator\footnote{Usually $X$ is the second quantisation of an essentially self-adjoint operator on $\mathcal{H}_+^{(1)}$, cf.\ \cite{Reed+} Section
VIII.10, Ex.\ 2.} with domain $\mathcal{D}$, which is the algebraic direct sum $\mathcal{D}:=\bigoplus_{n=0}^{\infty}\mathcal{D}_n$.

If $Y$ is a symmetric operator with domain $\mathcal{D}$ and $c,d\ge 0$ are constants satisfying
\begin{eqnarray}
\|Y\psi\|\le d\|N\psi\|+d\|\psi\|,\quad
\mathrm{Re}(\langle N\psi,(X+Y)\psi\rangle)\ge -c\langle\psi,(N+I)\psi\rangle
\end{eqnarray}
for all $\psi\in\mathcal{D}$, then $X+Y$ is essentially self-adjoint on $\mathcal{D}$.
\end{theorem}
Recall that positivity of $X_n$ means that $\langle\psi,X_n\psi\rangle\ge 0$ for all $\psi\in\mathcal{D}_n$.
\begin{proof*}
First note that $X$ is positive, because the $X_n$ are. Furthermore, by construction, $XN=NX=\sqrt{N}X\sqrt{N}\ge 0$ on $\mathcal{D}$ and hence also $(X+dN)^2\ge d^2N^2$.

To apply Konrady's trick we first note that the operator $X+dN$ is essentially self-adjoint on $\mathcal{D}$, because all $X_n+dnI$ are essentially self-adjoint on $\mathcal{D}_n\subset\mathcal{H}_+^{(n)}$. By the assumption and the previous estimate we then find for all $\psi\in\mathcal{D}$ that
\[
\|Y\psi\|\le d\|N\psi\|+d\|\psi\|\le \|(X+dN)\psi\|+d\|\psi\|.
\]
By W\"ust's Theorem \ref{Thm_Wuest} the operator $X+Y+dN$ is then essentially self-adjoint on $\mathcal{D}$. The point of Konrady's trick is now the following estimate, which follows from the assumptions:
\[
\|(X+Y+dN)\psi\|^2\ge\|(X+Y)\psi\|^2+\|dN\psi\|^2-2cd\langle\psi,(N+I)\psi\rangle
\ge\left(\|dN\psi\|-c\|\psi\|\right)^2-(c^2+2cd)\|\psi\|^2
\]
for all $\psi\in\mathcal{D}$. Thus we see by an elementary estimate that
\[
\|dN\psi\|\le \|(X+Y+dN)\psi\|+c'\|\psi\|
\]
for $c'=c+\sqrt{c^2+2cd}$ and another application of W\"ust's Theorem \ref{Thm_Wuest} now proves that $X+Y=(X+Y+dN)-dN$ is essentially self-adjoint on $\mathcal{D}$.
\end{proof*}
The usefulness of Theorem \ref{Thm_Konrady} lies in the fact that the required estimates may be easier to establish than the ones needed for W\"ust's Theorem.

Finally we will prove a result that concerns limits of self-adjoint operators. This will be relevant in Section \ref{Sec_SA} for second order Wick polynomials, which can be described as a coincidence point limit of self-adjoint operators. The Monotone Graph Limit Theorem below does not allow us to conclude essential self-adjointness of the limit on a certain domain, but it does provide a self-adjoint extension for which we have some control over the spectral projections. First we need a lemma:
\begin{lemma}\label{Lem_InvIneq}
If $X_2\ge X_1\ge 0$ are self-adjoint operators with a common core $\mathcal{D}$, then
$(X_1+I)^{-1}\ge(X_2+I)^{-1}\ge 0$.
\end{lemma}
\begin{proof*}
By the Spectral Theorem (\cite{Reed+} Section VIII.3) the operators $S_i:=(X_i+I)^{\frac{1}{2}}$ are well-defined and injective on $\mathcal{D}$. In fact, the given domain is a core, because the domain of $\overline{S_i|_{\mathcal{D}}}$ contains the domain of $X_i$, which, in turn, is a core for $S_i$, again by the Spectral Theorem. As $0$ is not in the spectrum of $S_i$, its range on $\mathcal{D}$ is dense in $\mathcal{H}$. Now let $Y:=S_1S_2^{-1}$ on $S_2\mathcal{D}$. Then $\|Y\psi\|\le\|\psi\|$,
so $Y$ is densely defined and bounded. It follows that $0\le\overline{Y}Y^*\le I$, where $Y^*=\overline{S_2^{-1}S_1}$.
This implies that $0\le (X_2+I)^{-1}\le (X_1+I)^{-1}$ on $S_1\mathcal{D}$ and hence on all of $\mathcal{H}$.
\end{proof*}

\begin{theorem}[Monotone Graph Limit Theorem]\label{Thm_graph}
Let $X_n$ be a sequence of self-adjoint operators on a Hilbert space $\mathcal{H}$ with a common core $\mathcal{D}$ and assume that $X_{n+1}\ge X_n\ge 0$ on $\mathcal{D}$ for all $n$. If the $X_n$ converge strongly to an operator $X$
on $\mathcal{D}$, then there is a self-adjoint extension $Y$ of $X$ such that $X_n\rightarrow Y$ in the strong resolvent sense, i.e.\ $(X_n+\lambda I)^{-1}$ converges strongly to $(Y+\lambda I)^{-1}$ for each $\lambda\in\mathbb{C}\setminus\mathbb{R}$.
\end{theorem}
Convergence in the strong resolvent sense entails that the spectral projections of $X_n$ converge strongly to those of $Y$ for all intervals $(a,b)\subset\mathbb{R}$, as long as $a$ and $b$ are not in the pure point spectrum of $Y$
(\cite{Reed+} I, Theorem VIII.24).
\begin{proof*}
We let $R_n:=(X_n+I)^{-1}$, so that $R_n\ge R_{n+1}\ge 0$ by Lemma \ref{Lem_InvIneq}. It follows from Lemma 5.1.4 in
\cite{Kadison+} that there is a positive bounded operator $R$ such that $R_n\rightarrow R$ strongly. Because of this and the strong convergence $X_n\rightarrow X$ on $\mathcal{D}$ we can apply the Graph Limit Theorem X.65 of \cite{Reed+} to find a self-adjoint operator $Y$ with $R=(Y+I)^{-1}$. Thus, $X_n\rightarrow Y$ in the strong resolvent sense. By Theorem VIII.26 loc.cit.\ the operator $Y$ is the strong graph limit of the $X_n$, which is an extension of $X$. We refer to \cite{Reed+} for a definition of the strong graph limit and only note that $Y$ may be a proper extension of the closure $\overline{X}$, so $X$ need not be essentially self-adjoint (see also the example below the Trotter-Kato Theorem VIII.22 loc.cit.).
\end{proof*}

\section{Quasi-free Hadamard states for a scalar quantum field}\label{Sec_QFHad}

We now want to consider a general framework for quantum field theory on a curved spacetime and the Fock spaces that are associated to quasi-free states. For this purpose we fix a (globally hyperbolic) spacetime $M=(\mathcal{M},g)$, where $\mathcal{M}$ is a spacetime manifold of arbitrary dimension $d\ge 2$ and $g$ a Lorentzian metric. We define the tensor algebra
\[
\alg{U}_M:=\bigoplus_{n=0}^{\infty}\Test_0(M)^{\otimes n},
\]
where we take the algebraic direct sum and tensor product and we set $\Test_0(M)^{\otimes 0}:=\C$. This algebra is the polynomial $^*$-algebra generated by an identity $I$ and the symbols $\Phi(f)$ with $f\in\Test_0(M)$, where $f\mapsto\Phi(f)$ is the obvious complex linear map into the summand $\Test_0(M)^{\otimes 1}$ of degree 1, and the $^*$-operation is defined by anti-linear extension of $\Phi(f)^*:=\Phi(\bar{f})$. Given an algebraic state $\omega$ on $\alg{U}_M$ we can construct the associated GNS-representation $\pi_{\omega}$ on the Hilbert space $\mathcal{H}_{\omega}$ with cyclic vector $\Omega_{\omega}$ and dense invariant domain $\mathcal{D}_{\omega}:=\pi_{\omega}(\alg{U}_M)\Omega_{\omega}$.

Let $\omega_2\in\mathcal{D}'(M^{\times 2})$ be a distribution which is of positive type, in the sense that $\omega_2(\overline{f},f)\ge 0$ whenever $f\in\Test_0(M)$. Notice that $2\omega_2$ is a semi-definite inner product on $\Test_0(M)$, so we may obtain a Hilbert space $\mathcal{K}_{2\omega_2}$ by dividing out the null-space of test-functions $f$ such that $2\omega_2(\bar{f},f)=0$ and then taking the Hilbert space completion.\footnote{The same Hilbert space is obtained in a different way in appendix A of \cite{Kay+1991}. Equality follows from Lemma A.1 loc.cit.} We denote the canonical projection map $C_0^{\infty}(M)\rightarrow\mathcal{K}_{2\omega_2}$ by $\kappa$ and we warn the reader that complex conjugation on $C_0^{\infty}(M)$ generally does not descend to a well-defined ($\mathbb{R}$-linear) map on the subspace $\kappa(C_0^{\infty}(M))$ of $\mathcal{K}_{2\omega_2}$, because $\kappa(f)=0$ does not necessarily imply that $\kappa(\overline{f})=0$. We also introduce the map
$\map{\overline{\kappa}}{C_0^{\infty}(M)}{\overline{\mathcal{K}_{2\omega_2}}}$ defined by $\overline{\kappa}(f):=j^{-1}(\kappa(\bar{f}))$, where $\map{j}{\overline{\mathcal{K}_{\omega_2}}}{\mathcal{K}_{\omega_2}}$
is the anti-linear bijection. Note that $\overline{\kappa}$ is linear (and not anti-linear) and that
\begin{equation}\label{Eqn_cc}
\langle\overline{\kappa}(f),\overline{\kappa}(g)\rangle=\overline{\omega}_2(\overline{f},g).
\end{equation}

For the Hilbert space $\mathcal{K}_{2\omega_2}$ we can construct the symmetric Fock space $\mathcal{H}_+$ as in Section \ref{Sec_Fock}. For convenience we define
\[
\alpha_+^*(f):=a_+^*(\kappa(f)),\quad \alpha_+(f):=a_+(j(\overline{\kappa}(f)))
\]
for each $f\in\Test_0(M)$ and we set furthermore
\begin{equation}\label{Eqn_Field}
\Phi'(f):=\frac{1}{\sqrt{2}}(\alpha_+(f)+\alpha_+^*(f))
\end{equation}
as a densely defined operator on the subspace $\mathcal{F}_+$ of $\mathcal{H}_+$. Note that $\Phi'(f)$ is complex linear in $f$.

Let $\mathcal{D}$ denote the linear space generated by polynomials in the operators $\Phi'(f)$ acting on the Fock vacuum vector $\Omega$. Then one has (see also \cite{Kay+1991} Section 3.2)
\begin{proposition}
There is a unique quasi-free state $\omega$ on $\alg{U}_M$ with the given two-point distribution $\omega_2$ and its GNS-representation $(\mathcal{H}_{\omega},\pi_{\omega},\mathcal{D}_{\omega},\Omega_{\omega})$ is given by $\mathcal{H}_{\omega}:=\mathcal{H}_+$, $\Omega_{\omega}:=\Omega$, $\mathcal{D}_{\omega}:=\mathcal{D}$ and $\pi_{\omega}(\Phi(f)):=\Phi'(f)|_{\mathcal{D}}$.
\end{proposition}
\begin{proof*}
The equality $\langle \Omega,\pi_{\omega}(A)\Omega\rangle=\omega(A)$ where $\omega$ is the quasi-free state with the given $\omega_2$ follows from the commutation relations between creation and annihilation operators. (This also proves the existence of this quasi-free state.) The other properties which characterise the GNS-representation can be inferred\footnote{Notice that this reference uses the Segal field $\Phi_S(u):=\frac{1}{\sqrt{2}}(a_+^*(u)+a_+(u))$, which is only real linear in $u\in\mathcal{K}_{\omega_2}$, but has the advantage that it can be defined on any abstract Fock space, since no conjugation of $u$ is needed. To check e.g.\ cyclicity for the physical field $\Phi(f)$ one uses that for real-valued $f$ we have $\Phi_S(\kappa(f))=\Phi(f)$ and $\Phi_S(i\kappa(f))=\Phi(if)-i\sqrt{2}\alpha_+(f)$, which allows us to approximate any vector of the form $\Phi_S(\kappa(f)_1)\cdots\Phi_S(\kappa(f)_n)\Omega$ by $\Phi(f_1)\cdots\Phi(f_n)\Omega$ plus terms with $<n$ particles, which can be dealt with by induction. } e.g.\ from Proposition 5.2.3 in \cite{Bratteli+}.
\end{proof*}

For any vector $\psi\in\mathcal{D}_{\omega}$ one can now show that $(f,g)\mapsto\langle\psi,\Phi'(f)\Phi'(g)\psi\rangle$ is a distribution on $M^{\times 2}$, which means that $f\mapsto\Phi'(f)\psi$ is an $\mathcal{H}_{\omega}$-valued distribution (cf.\ \cite{Strohmaier+2002}). In order to describe Wick polynomials (including derivatives) in terms of Hilbert space-valued distributions we first need to recall some notions from microlocal analysis (see \cite{Hoermander}).

To any distribution $u$ on $M$ one can associate the wave front set $WF(u)\subset T^*M$, which describes the singularities of $u$. A point $(x,k)\in WF(u)$ should be thought of as a singularity at $x$ in the direction $k\not=0$. (The wave front set is invariant under positive re-scaling of $k$). If $\mathcal{Z}$ denotes the zero section of $T^*M$ and $\Gamma\subset T^*M\setminus\mathcal{Z}$ is a closed conic subset, i.e.\ if $\Gamma$ is invariant under positive re-scaling at every point in $M$, then we denote by $\mathcal{D}'_{\Gamma}(M)$ the linear space of distributions whose wave front set is contained in $\Gamma$. This space can be endowed with a notion of convergence for sequences, known as the H\"ormander pseudo-topology (we refer to \cite{Hoermander} for its detailed definition). The space of test-functions $C_0^{\infty}(M)$ is dense in $\mathcal{D}'_{\Gamma}(M)$ in this pseudo-topology. Given two closed conic subsets $\Gamma,\Gamma'\subset T^*M\setminus\mathcal{Z}$ such that $\Gamma\cap -\Gamma'=\emptyset$ (where $-$ denotes the multiplication of vectors by $-1$, defined pointwise in the cotangent space) one can uniquely extend the (pointwise) product from $C_0^{\infty}(M)\times C_0^{\infty}(M)$ to $\mathcal{D}'_{\Gamma}(M)\times\mathcal{D}'_{\Gamma'}(M)$ in a sequentially continuous way. This result is known as H\"ormander's criterion for the multiplication of distributions.

Now we make an additional assumption on $\omega$, namely that it satisfies the microlocal spectrum condition \cite{Brunetti+3}. For a quasi-free state this means that $\omega_2$ is of (generalised) Hadamard form \cite{Radzikowski1996,Sanders}, i.e.\ that
\begin{equation}\label{Eqn_Had}
WF(\omega_2)\subset (V^-\setminus \mathcal{Z})\times (V^+\setminus \mathcal{Z}),
\end{equation}
where $V^{\pm}$ are the fiber bundles of future ($+$) and past ($-$) pointing causal covectors on the spacetime $M$. The free scalar field is known to have quasi-free Hadamard states in any globally hyperbolic spacetime \cite{Fulling+}. The microlocal spectrum condition allows one to prove the following
\begin{lemma}\label{Lem_Wick}
For each $l,m$, the $\mathcal{K}_{2\omega_2}^{\otimes l}\otimes\overline{\mathcal{K}}_{2\omega_2}^{\otimes m}$-valued distribution
\[
(\kappa^{\otimes l}\otimes\overline{\kappa}^{\otimes m})(f_1,\ldots,f_l,h_1,\ldots,h_m):=
\kappa(f_1)\otimes\ldots\otimes\kappa(f_l)\otimes\overline{\kappa}(h_1)\otimes\ldots\otimes\overline{\kappa}(h_m)
\]
on $M^{\times (l+m)}$ has a wave front set contained in $(V^+)^{\times l}\times (V^-)^{\times m}$ and it can be extended in a unique way to $\mathcal{D}'_{\Gamma}(M^{\times(l+m)})$, if
$\Gamma\cap((V^+)^{\times l}\times (V^-)^{\times m})=\emptyset$, using the H\"ormander pseudo-topology.
\end{lemma}
The definition of the distribution $\kappa^{\otimes l}\otimes\overline{\kappa}^{\otimes m}$ on $M^{\times (l+m)}$ uses the Schwartz Kernel Theorem (c.f.\ Theorem 5.2.1 in \cite{Hoermander} for the scalar case). The estimates of the wave front sets of $\kappa^{\otimes l}$ and $\overline{\kappa}^{\otimes l}$ and the extension to $\mathcal{D}'_{\Gamma}$ follow from Equation (\ref{Eqn_cc}) and the results in Section 8.2 of \cite{Hoermander}, extended to the case of Hilbert space-valued distributions (using e.g. Proposition 2.2 in \cite{Strohmaier+2002} or Theorem A.1.3 in \cite{SandersT}).

Lemma \ref{Lem_Wick} can be combined with Proposition \ref{Prop_Nest} in order to smear polynomials in $\alpha^*_+$ and $\alpha_+$ with certain distributions which still yield well-defined operators on $\mathcal{F}_+\subset\mathcal{H}_{\omega}$. However, an even stronger estimate on the wave front set can be obtained by exploiting the smaller domain $\mathcal{D}_{\omega}$:
\begin{theorem}\label{Thm_Wick}
For each $\psi\in\mathcal{D}_{\omega}$ and $l,m$, the Hilbert space-valued distribution
\[
A^{(l,m)}_{\psi}(f_1,\ldots,f_l,h_1,\ldots,h_m):=\alpha^*_+(f_1)\cdots\alpha^*_+(f_l)\alpha_+(h_1)\cdots\alpha_+(h_m)\psi
\]
on $M^{\times (l+m)}$ has a wave front set contained in $(V^+)^{\times l}\times \mathcal{Z}^{\times m}$. It can be extended in a unique way to $\mathcal{D}'_{\Gamma}(M^{\times (l+m)})$, if
$\Gamma\cap\left((V^+)^{\times (l+m)}\cup (V^-)^{\times (l+m)}\right)=\emptyset$, using the H\"ormander pseudo-topology.
\end{theorem}
Here we follow the convention to exclude both future and past pointing vectors from $\Gamma$, to ensure that the commutation relations still make sense (see \cite{Duetsch+2001,Hollands+2003}).
\begin{proof*}
First we consider the case $l=0$. The wave front set of the $\mathcal{H}_{\omega}$-valued distribution
$A^{(0,m)}_{\psi}$ can be estimated by considering
\[
w(h'_1,\ldots,h'_m,h_1,\ldots,h_m):=\langle A^{(0,m)}_{\psi}(\overline{h'}_1,\ldots,\overline{h'}_m),
A^{(0,m)}_{\psi}(h_1,\ldots,h_m)\rangle.
\]
Writing $\psi=\sum_{i=0}^N\psi_i\in\mathcal{D}_{\omega}$ with $\psi_0\in\mathbb{C}$ and representing $\psi_i$ as a symmetric function in $\Test_0(M^{\times i})$ for $i>0$ we can see from the definition of the creation and annihilation operators that $w$ is smooth, because it is a finite sum of terms of the form
\begin{eqnarray}
&&\int_{M^{\times 2n}}\overline{\psi_{m+n}}(x'_1,\ldots,x'_{m+n})\psi_{m+n}(y'_1,\ldots,y'_{m+n})
\prod_{j=1}^m\omega_2(x'_j,x_j)\omega_2(y_j,y'_j)\prod_{j=m+1}^{m+n}\omega_2(x'_j,y'_j),\nonumber
\end{eqnarray}
where we omitted the metric induced volume forms on $M$. These are smooth functions of $x_i,y_i$, $1\le i\le m$ because of Theorem 8.2.12 in \cite{Hoermander} and the fact that $(x,k;y,l)\in WF(\omega_2)$ implies both $k\not=0$ and $l\not=0$. Now, by Proposition \ref{Prop_Nest} we have
\[
\|A^{(l,m)}_{\psi}(f_1,\ldots,f_l,h_1,\ldots,h_m)\|\le C\prod_{i=1}^l\|\kappa(f_i)\|\cdot
\|A^{(0,m)}_{\psi}(h_1,\ldots,h_m)\|
\]
for some $C>0$ that may depend on $\psi$. Hence,
\[
WF(A^{(l,m)}_{\psi})\subset WF(\kappa^{\otimes l}\otimes A^{(0,m)}_{\psi})\subset
(WF(\kappa)\cup\mathcal{Z})^{\times l}\times(WF(A^{(0,m)}_{\psi})\cup\mathcal{Z}^{\times m})
\subset (V^+)^{\times l}\times\mathcal{Z}^{\times m}.
\]
The extension to $\mathcal{D}'_{\Gamma}$ again follows from the results in Section 8.2 of \cite{Hoermander}, extended to the case of Hilbert space-valued distributions.
\end{proof*}

For each $\psi\in\mathcal{D}_{\omega}$ one now defines $\pi_{\omega}(\Phi^{\otimes n})\psi$ as the $\mathcal{H}_{\omega}$-valued distribution on $M^{\times n}$ obtained by expanding $\Phi$ in terms of creation and annihilation operators and the normally ordered tensor product $:\pi_{\omega}(\Phi^{\otimes n}):\psi$ is obtained by moving all creation operators in the expression for $\pi_{\omega}(\Phi^{\otimes n})\psi$ to the left of all annihilation operators. Wick polynomials can now be defined by restricting the distribution $:\pi_{\omega}(\Phi^{\otimes n}):\psi$ on $M^{\times n}$ to the diagonal $\Delta:=\left\{(x,\ldots,x)\in M^{\times n}|\ x\in M\right\}$ as follows. Let $\delta^{(n)}\in\mathcal{D}'(M^{\times n})$ be the distribution $\delta^{(n)}(h):=\int_Mh|_{\Delta}$. Given any
$\psi\in\mathcal{D}_{\omega}$, $f\in\Test_0(M)$ and any partial differential operator $Q$ on $M^{\times n}$ with smooth coefficients and formal adjoint $Q^*$ we consider the expression
\[
:\pi_{\omega}(Q^*\Phi^n):(f)\psi\ :=\ :\pi_{\omega}(\Phi^{\otimes n}):(Q(f\delta^{(n)}))\psi.
\]
If $Q=1$ this expression is called the $n^{\mathrm{th}}$ Wick-power.
\begin{corollary}\label{Cor_Wick}
Given $f$ and $Q$, the operator $:\pi_{\omega}(Q^*\Phi^n):(f)$ is well-defined on $\mathcal{D}_{\omega}$ by Theorem \ref{Thm_Wick} and it is a strong limit on $\mathcal{D}_{\omega}$ of operators in $\alg{U}_M$. For each $\psi\in\mathcal{D}_{\omega}$ the map $f\mapsto :\pi_{\omega}(Q^*\Phi^n):(f)\psi$ is an $\mathcal{H}_{\omega}$-valued distribution on $M$.
\end{corollary}
\begin{proof*}
Set $\Gamma:=\left\{(x,k_1;\ldots;x,k_n)|\ \sum_{j=1}^nk_j=0\right\}=WF(\delta^{(n)})$. For every partial differential operator $Q$ on $M^{\times n}$ the map $f\mapsto Q(f\delta^{(n)})$ is sequentially continuous from $C_0^{\infty}(M)$ to $\mathcal{D}'_{\Gamma}(M^{\times n})$. By Theorem \ref{Thm_Wick} there are well-defined normally ordered operators $A^{(l,n-l)}(Q(f\delta^{(n)}))$ defined for $\psi\in\mathcal{D}_{\omega}$ by
$A^{(l,n-l)}(Q(f\delta^{(n)}))\psi:=A^{(l,n-l)}_{\psi}(Q(f\delta^{(n)}))$. For each $\psi\in\mathcal{D}_{\omega}$ this operator defines a $\mathcal{H}_{\omega}$-valued distribution and the Wick polynomials are just finite sums of these distributions.
\end{proof*}

All Hadamard two-point distributions of a free scalar field with a fixed mass are known to differ pairwise by a smooth function. Moreover, their singularity structure is locally determined by the geometry of the spacetime alone. These remarks allow a more advanced formulation of the normal ordering procedure and the construction of the Wick powers than what we have given here, which is local, covariant and representation independent (cf.\ \cite{Brunetti+,Hollands+2003}). Note that such a different notion of normal ordering would not influence the results of Section \ref{Sec_SA} below, because the difference would be a finite multiple of the identity operator.

\section{Self-adjointness of second order Wick polynomials}\label{Sec_SA}

In this section we present and prove our main results concerning the (essential) self-adjointness of second order Wick polynomials. We fix a spacetime $M$ and a quasi-free Hadamard state $\omega_2$ and consider the operator
\begin{equation}\label{Eq_DefT}
T=\sum_{j\in J}:\pi_{\omega}(\Phi^2):(Q^*_j\otimes Q_j(f_j\cdot\delta^{(2)}))
\end{equation}
with a finite index set $J$ and partial differential operators $Q_j$ on $M$ with real, smooth coefficients and
$f_j\in C_0^{\infty}(M)$. We let $K\subset M$ denote the union of the supports of the test-functions $f_j$. Furthermore we let
\[
T_1:=P_{+,1}TP_{+,1}
\]
be the compression of $T$ to the one-particle Hilbert space $\mathcal{H}^{(1)}\simeq\mathcal{K}_{2\omega_2}$.

We start by establishing the locality properties of the self-adjointness problem. For any open set $O\subset M$ we let
$\mathcal{D}_{\omega}(O)$ denote the subspace of $\mathcal{H}_{\omega}$ generated by polynomials in the fields smeared with test-functions supported in $O$ and we set $\mathcal{H}_{\omega}(O):=\overline{\mathcal{D}_{\omega}(O)}$. Then we have the
\begin{lemma}[Locality Lemma]\label{Lem_loc}
If $O\subset M$ is a relatively compact open set containing $K$, then $T$ restricts to a densely defined operator on $\mathcal{D}_{\omega}(O)\subset\mathcal{H}_{\omega}(O)$. If for every relatively compact open set $O\subset M$ containing $K$ the restriction of $T$ to $\mathcal{H}_{\omega}(O)$ is essentially self-adjoint on $\mathcal{D}_{\omega}(O)$, then $T$ is essentially self-adjoint on $\mathcal{D}_{\omega}$.
\end{lemma}
\begin{proof*}
If $O$ contains $K$, then the range of $T$ is contained in $\mathcal{H}_{\omega}(O)$, because
$(Q^*_j\otimes Q_j(f_j\cdot\delta^{(2)}))$ can be approximated by tensor products of test-functions supported in $O$. This proves the first statement. Now suppose that arbitrary $\psi\in\mathcal{H}_{\omega}$ and $\epsilon>0$ are given. We may find an operator $A\in\alg{U}_M$ such that $\|\psi-\pi_{\omega}(A)\Omega_{\omega}\|<\frac{\epsilon}{2}$. Now choose $O$ to be a relatively compact neighbourhood of $K$ and of the supports of all test-functions occurring in $A$. If we may choose $O$ in such a way that the restriction of $T$ to $\mathcal{H}_{\omega}(O)$ is essentially self-adjoint, then we may find $A_{\pm}\in\alg{U}_M$ with test-functions supported in $O$ such that
$\|(\pi_{\omega}(A)-(T\pm iI)\pi_{\omega}(A_{\pm}))\Omega_{\omega}\|<\frac{\epsilon}{2}$. Now
$\|\psi-(T\pm iI)\pi_{\omega}(A_{\pm}))\Omega_{\omega}\|<\epsilon$, so the range of $T\pm iI$ is dense for both signs and therefore $T$ is essentially self-adjoint.
\end{proof*}
A similar statement holds at the level of one-particle Hilbert spaces for $T_1$.

If we write out $T$ in terms of creation and annihilation operators, then the terms which consist of two creation operators or two annihilation operators are not problematic. Indeed, they yield an essentially self-adjoint operator by Proposition \ref{Prop_NelsonFock}, combined with Lemma \ref{Lem_Wick} and the fact that
\[
WF((Q_j^*\otimes Q_j)(f_j\delta^{(2)}))\subset WF(\delta^{(2)})=\left\{(x,k;x,-k)\right\}.
\]
The term consisting of a creation and annihilation operator, however, cannot be dealt with in the same way, unless the singularities in $\omega_2$ are very mild. Indeed, this term is the second quantisation of its compression $T_1$ to $\mathcal{K}_{2\omega_2}$, which is in general not a Hilbert-Schmidt operator. When $T_1$ is nevertheless a bounded operator one can still prove an analogue of Proposition \ref{Prop_NelsonFock}, but this too cannot be expected unless the singularities in $\omega_2$ are not too strong.

Because our aim is to prove results which are as general as possible, we will allow the possibility that $T_1$ is unbounded, so we need to use different methods to prove the essential self-adjointness of $T$. For this purpose we consider the case that the $f_j$ are in the following set of functions.
\begin{definition}
$\mathcal{S}$ is the set of $f\in C_0^{\infty}(M)$ which can be written as a finite sum of squares of real-valued test-functions.
\end{definition}
Note in particular that $f\in\mathcal{S}$ entails $f\ge 0$. When $f_j\in\mathcal{S}$ for all $j$ it is known that $T$ is semi-bounded \cite{Fewster+2002_3}, which implies that $T$ certainly has self-adjoint extensions (e.g.\ the Friedrichs extension). In Theorem \ref{Thm_SA} below we prove moreover that $T$ is essentially self-adjoint when $T_1$ is, but first we prove that we can always pick out a particular self-adjoint extension of $T$ in a nice way.
\begin{theorem}\label{Thm_SAGL}
Assume that $f_j\in\mathcal{S}$ for all $f_j$ in Equation (\ref{Eq_DefT}). Then we may choose a sequence
$F_n\in C_0^{\infty}(M^{\times 2})$ and a sequence $c_n\in\mathbb{R}_{\ge 0}$ such that for each quasi-free state $\omega$
\begin{enumerate}
\item the operators $X_n:=\ :\pi_{\omega}(\Phi^{\otimes 2}):(F_n)+c_nI$ are essentially self-adjoint on
$\mathcal{D}_{\omega}$,
\item $c_n$ converges to some $c\in\mathbb{R}$ and $X_n$ converges strongly to $T+cI$ on $\mathcal{D}_{\omega}$,
and
\item $X_{n+1}\ge X_n\ge 0$ for all $n$.
\end{enumerate}
Consequently, there is a self-adjoint extension $X$ of $T$ such that $X_n\rightarrow X$ in the strong resolvent sense.
\end{theorem}
\begin{proof*}
By rearranging sums we may assume without loss of generality that $f_j=\tilde{f}_j^2$ with
$\tilde{f}_j\in C_0^{\infty}(M,\mathbb{R})$. In the following we will say that a coordinate neighbourhood $U$ is ''suitably small'' if we can choose the coordinates $x$ such that one of them, $x^0$, is a time-coordinate. Recall that the compact set $K$ is the union of the supports of all $\tilde{f}_j$. We can cover $K$ by a finite number of suitably small coordinate neighbourhoods $O_k$, $1\le k\le N$, which we augment by $O_0:=M\setminus K$. Let
$\left\{\phi_k\right\}_{k=0,\ldots,N}$ be a partition of unity subject to the cover $\left\{O_k\right\}_{k=0,\ldots,N}$, i.e.\ $\phi_k\in C_0^{\infty}(O_k)$ and $\phi_k\ge 0$ for all $k$ and $\sum_{k=0}^N\phi_k\equiv 1$ on $M$. Now set\footnote{This choice of partition of unity is not new, cf.\ \cite{Hoermander1966} p.142 for a similar construction. The existence of a partition of unity consisting of squares of test-functions is rather remarkable in the light of Remark \ref{Rem_SSQ} below.}
\[
\chi_k:=\phi_k\left(\sum_{k=0}^N\phi_k^2\right)^{-\frac{1}{2}}.
\]
Notice that $\chi_k\in C_0^{\infty}(O_k)$ and $\sum_{k=0}^N\chi_k^2\equiv 1$. Since $\chi_0\tilde{f}_j\equiv 0$ we can consider the partition $\tilde{f}_j=\sum_{k=1}^N\tilde{f}_{j,k}$ with $\tilde{f}_{j,k}:=\tilde{f}_j\chi_k^2$, so that $\tilde{f}_j^2=\sum_{k,l=1}^N\tilde{f}_{j,k}\tilde{f}_{j,l}$. Each term in the latter sum is of the form $\tilde{f}_{j,k}\tilde{f}_{j,l}=(\tilde{f}_j\chi_k\chi_l)^2$ with $\tilde{f}_j\chi_k\chi_l$ real-valued and supported in a suitably small coordinate neighbourhood. After relabeling we may therefore assume without loss of generality that all $\tilde{f}_j$ are supported in suitably small neighbourhoods.

We partition the set $J$ into disjoint subsets $J_k$, such that $j\in J_k$ implies
$\mathrm{supp}\tilde{f}_j\subset O_k$. Taking advantage of the special coordinates in the suitably small neighbourhood $O_k$ we define
\begin{eqnarray}
F_{n,k}(x,y)&:=&(2\pi)^{-4}\sum_{j\in J_k}\int_{B_n} dp\ Q_j(\tilde{f}_j(x)e^{-ip\cdot x})Q_j(\tilde{f}_j(y)
e^{ip\cdot y})\nonumber\\
F'_{n,k}(x,y)&:=&2(2\pi)^{-4}\sum_{j\in J_k}\int_{B'_n} dp\ Q_j(\tilde{f}_j(x)e^{-ip\cdot x})
Q_j(\tilde{f}_j(y)e^{ip\cdot y}),\nonumber
\end{eqnarray}
where $B_n:=\left\{p|\ |p_{\mu}|\le n\right\}$ is a cube and $B'_n:=B_n\cap\left\{p_0\ge 0\right\}$. Then we define $F_n:=\sum_{k=1}^NF_{n,k}$ and $F'_n:=\sum_{k=1}^NF_{n,k}$. Because each $B_n$ is bounded the $F_{n,k}$ and $F'_{n,k}$ are smooth functions and therefore the $F_n$ and $F'_n$ are smooth too. We set $c_n:=\omega_2(F'_n)$ and it now remains to check the properties of $X_n:=\ :\pi_{\omega}(\Phi^{\otimes 2}):(F_n)+c_nI$.

By Lemma \ref{Lem_Wick} and Proposition \ref{Prop_NelsonFock} we see that the $X_n$ have $\mathcal{D}_{\omega}$ as a dense set of analytic vectors in the GNS-representation of the quasi-free state $\omega$. They are also symmetric and therefore essentially self-adjoint on $\mathcal{D}_{\omega}$. Furthermore, as $n\rightarrow\infty$ the functions $F_n$ converge to $F$ in the H\"ormander pseudo-topology, so by Theorem \ref{Thm_Wick} the operators $X_n-c_nI$ converge strongly on $\mathcal{D}_{\omega}$ to $T$. Next we note that the $c_n$ are non-negative, because
\begin{equation}\label{Eqn_positivity}
\omega_2(F'_{n,k})=2(2\pi)^{-4}\sum_{j\in J_k}\int_{B'_n} dp\
\omega_2\left(\overline{Q_j(\tilde{f}_je^{ip\cdot .})},Q_j(\tilde{f}_je^{ip\cdot .})\right)\ge 0.
\end{equation}
In fact, using the Hadamard condition and the special properties of the coordinates used to define the $F'_{n,k}$ and the fact that $B'_n$ is a half-space one can show that $c:=\lim_{n\rightarrow\infty}c_n$ is finite. Finally, each $\psi\in\mathcal{D}_{\omega}$ with $\|\psi\|=1$ defines a distribution $\omega'_2(f,h):=\langle\psi,\pi_{\omega}(\Phi(f)\Phi(h))\psi\rangle$ with $\omega'_2(\overline{f},f)\ge 0$ and hence
\[
\langle\psi,X_n\psi\rangle=(\omega_2'-\omega_2)(F_n)+c_n=(\omega_2'-\omega_2)(F'_n)+\omega_2(F'_n)=\omega_2'(F'_n),
\]
where we used the symmetry of $(\omega'_2-\omega_2)(x,y)$ in the second equality. As in equation (\ref{Eqn_positivity}), we now find that $X_{n+1}\ge X_n$ for all $n$ and that $X_1\ge 0$. The final claim now follows from the Monotone Graph Limit Theorem \ref{Thm_graph}.
\end{proof*}

We will now prove that $T$ is essentially self-adjoint when the compression $T_1$ is essentially self-adjoint, using Konrady's trick.
\begin{theorem}\label{Thm_SA}
Assume that $f_j\in\mathcal{S}$ for all $f_j$ in Equation (\ref{Eq_DefT}). If $T_1$ is essentially self-adjoint, then so is $T$.
\end{theorem}
\begin{proof*}
Let $F:=\sum_{j\in J}Q_j\otimes Q_j(\tilde{f}_j(x)\tilde{f}_j(y)\delta^{(2)}(x,y))$ and define the operators
$X:=\alpha_+^*\otimes\alpha_+(F)$ and $Y:=\frac{1}{2}((\alpha_+^*)^{\otimes 2}+\alpha_+^{\otimes 2})(F)$ on $\mathcal{D}_{\omega}$ (cf.\ Corollary \ref{Cor_Wick}), so that $T=\ :\Phi^2:(F)=X+Y$. Our aim is to apply Konrady's trick in the form of Theorem \ref{Thm_Konrady}.

First note that $T_1=P_{+,1}XP_{+,1}$ is essentially self-adjoint and positive on the dense domain
$\mathcal{D}_{\omega}\cap\mathcal{K}_{2\omega_2}=\mathcal{H}_+^{(1)}$. It then easily follows that $X$ is essentially self-adjoint and positive on $\mathcal{D}_{\omega}$, because it is the second quantisation of $P_{+,1}XP_{+,1}$
(cf.\ \cite{Reed+} Section VIII.10, Ex.\ 2.). By Lemma \ref{Lem_Wick} and Proposition \ref{Prop_Nest} the operator $Y$ satisfies $\|Y\psi\|\le \frac{d}{2}\|(N+2)\psi\|\le d\|N\psi\|+d\|\psi\|$ for some $d\ge 0$. To obtain the final estimate needed to apply Theorem \ref{Thm_Konrady} we note that for any $h\in C_0^{\infty}(M)$ we have
\[
N\pi_{\omega}(\Phi(h))=\pi_{\omega}(\Phi(h))(N-I)+\sqrt{2}\alpha_+^*(h).
\]
Because $:\pi_{\omega}(\Phi^{\otimes 2}):(\overline{h},h)
=\pi_{\omega}(\Phi(\overline{h}))\pi_{\omega}(\Phi(h))-\omega_2(\overline{h},h)I$ it follows that
\begin{eqnarray}
&&N:\pi_{\omega}(\Phi(\overline{h}))\pi_{\omega}(\Phi(h)):+
:\pi_{\omega}(\Phi(\overline{h}))\pi_{\omega}(\Phi(h)):N\nonumber\\
&=&
2\pi_{\omega}(\Phi(\overline{h}))N\pi_{\omega}(\Phi(h))+\alpha_+^*(\overline{h})\alpha_+(h)
-\alpha_+^*(h)\alpha_+(\overline{h})-2\omega_2(\overline{h},h)(N+I)\nonumber\\
&\ge&-4\omega_2(\overline{h},h)(N+I),\nonumber
\end{eqnarray}
where we used the elementary estimates above Proposition \ref{Prop_Nest} in the final inequality.

Referring to the proof of Theorem \ref{Thm_SAGL} we may assume without loss of generality that all $\tilde{f}_j$ are supported in a single, suitably small coordinate neighbourhood. We may then write
\[
Q_j\otimes Q_j(\tilde{f}_j(x)\tilde{f}_j(y)\delta^{(2)}(x,y))=(2\pi)^{-4}\int dk\ Q_j(\tilde{f}_j(x)
e^{-ik\cdot x})Q_j(\tilde{f}_j(y)e^{ik\cdot y})
\]
and exploit the symmetry of the Wick square to integrate over $k_0\ge 0$ only, which yields for any $\psi\in\mathcal{D}_{\omega}$
\begin{eqnarray}\label{Eqn_KonrLocEst}
&&\mathrm{Re}(\langle\psi,N:\pi_{\omega}(\Phi^2):(Q_j\otimes Q_j(\tilde{f}_j\otimes \tilde{f}_j\delta^{(2)}))\psi\rangle)\nonumber\\
&=&(2\pi)^{-4}\int dk\ \mathrm{Re}\left(\langle\psi,N:\pi_{\omega}(\Phi^{\otimes 2}):
\left(Q_j(\tilde{f}_je^{-ik\cdot .}),Q_j(\tilde{f}_je^{ik\cdot .})\right)\psi\rangle\right)\nonumber\\
&=&2(2\pi)^{-4}\int_{k_0\ge 0} dk\ \mathrm{Re}\left(\langle\psi,N:\pi_{\omega}(\Phi^{\otimes 2}):
\left(Q_j(\tilde{f}_je^{-ik\cdot .}),Q_j(\tilde{f}_je^{ik\cdot .})\right)\psi\rangle\right)\nonumber\\
&\ge&-4(2\pi)^{-4}\int_{k_0\ge 0} dk\ \omega_2\left(Q_j(\tilde{f}_je^{-ik\cdot .}),Q_j(\tilde{f}_je^{ik\cdot .})\right)
\|\sqrt{N+I}\psi\|^2\nonumber\\
&\ge&-c_j\|\sqrt{N+I}\psi\|^2,\nonumber
\end{eqnarray}
where $c_j<\infty$ in the final inequality, because of the Hadamard condition and the fact that $k_0\ge 0$. After summing over $j$ we have verified all the assumptions of Theorem \ref{Thm_Konrady}, which completes the proof.
\end{proof*}

It is in order to make a remark on the class $\mathcal{S}$ of smearing functions appearing in Theorems \ref{Thm_SAGL} and \ref{Thm_SA} (see also \cite{Fewster+2002_3} p.345 for similar comments):
\begin{remark}\label{Rem_SSQ}
\textnormal{Note that a sum of squares of real-valued test-functions is clearly a positive test-function, but the converse is not true because of Hilbert's Theorem of 1888 \cite{Rudin2000}. In fact, if $O\subset\mathbb{R}^d$ is an open set with $d\ge 4$ then one may find a homogeneous polynomial $P\ge 0$ on $O$ of degree $4$ which cannot be written as a finite sum of squares of polynomials, and therefore it is not a finite sum of squares of $C^2$ functions either \cite{Bony+2006}.\footnote{In view of this, the heart of the problem seems to be algebraic in nature rather than analytic (at least in higher dimensions). It therefore seems unrelated to the infinite order zeroes of $f$, which are mentioned in the context of the one dimensional case in \cite{Fewster+2002_3}.} (For $d=3$ the same argument gives a counter-example for finite sums of squares of $C^3$ functions, using a polynomial of degree $6$.) Multiplying $P$ by a test-function $f$ which is identically $1$ near $0$ does not spoil the argument, so $fP$ is not a square of $C^2$ functions either. More generally, if $f(0)\not=0$ then $f^2P$ cannot be a sum of squares of $C^2$ functions. The work of \cite{Blekherman2006} suggests that such counter-examples could be plentiful. As a positive result, however, any nonnegative smooth function $f\ge 0$ on $O$ can be written as a finite sum of $C^1$ functions with Lipschitz continuous derivatives \cite{Bony+2006}, but this small amount of regularity places severe restrictions on the order of the $Q_j$ and of $\omega_2$. The fact that any $f\ge 0$ can be written as a difference of two squares of smooth functions seems of little use for our proof.}

\textnormal{Thus, our class of smearing functions is smaller than the class of all non-negative smooth functions. However, for some choices of $T$, such as the components of the stress-energy-momentum tensor, one could argue that one is really interested in the smearing function $1$ and one only uses test-functions to avoid a divergence caused by an integration over all of spacetime. As shown in the proof of Theorem \ref{Thm_SAGL}, there exist partitions of unity consisting of squares of real-valued functions, which would then be sufficient for these purposes.}
\hfill$\oslash$\smallskip\par\ignorespaces
\end{remark}

The results above have only made use of the microlocal spectrum condition, but for the case of free fields more detailed information is available. In the remainder of this section we will study especially the Wick square of a free scalar field and use the results above to establish its self-adjointness. We start with a lemma and a proposition concerning the compression to the one-particle Hilbert space. In these results we make use of the Sobolev wave front set $WF_{(s)}$ of a distribution with $s\in\mathbb{R}$. For its definition and properties we refer to the literature \cite{Junker+,HoermanderNlin,Duistermaat+1972}.\footnote{Although \cite{Junker+} provides the best overview of this material, it omits proofs and we note that the bottom line of its Theorem B5 seems to be erroneous because of the counter-example $(0,0;0,1)\in WF_{(-1)}(\delta^{\otimes 2})\setminus
\left(WF_{(0)}(\delta)\times\left(WF(\delta)\cup\left\{(0,0)\right\}\right)\cup\left(WF(\delta)\cup
\left\{(0,0)\right\}\right)\times WF_{(-1)}\right)$.}
\begin{lemma}\label{Lem_WFs+}
Let $\omega_2$ be a Hadamard two-point distribution of a free scalar field and $v\in \mathcal{E}'(M)$. Assume
for some $s\in\mathbb{R}$ that $WF_{(s)}(v)=\emptyset$. Then $WF_{(s+1)}(\omega_2(v,.))=\emptyset$.
\end{lemma}
In fact one can even prove the slightly stronger result that $WF_{\left(s+\frac32\right)}(\omega_2(v,.))=\emptyset$ using Theorem B.9 and Equations (62,63) in \cite{Junker+}, but we will not need this strengthened version here.
\begin{proof*}
We consider the distribution $u:=\omega_2(v,.)=\omega_2(.,v)+iE(v,.)$, which is well-defined and has $WF(u)\subset V^+$ by the Hadamard condition (see \cite{Hoermander} Theorem 8.2.13). As the first term on the right-hand side has a wave front set contained in $V^-$ we see that $WF_{(s)}(u)=WF_{(s)}(E(v,.))\cap V^+$ for all $s$. Now suppose that $WF_{(s)}(v)=\emptyset$ and $(x,k)\in WF_{(s+1)}(u)$. Then we have $(x,k)\in WF_{(s+1)}(E^{\pm}(v,.))$ for at least one choice of the sign. Because $v=KE^{\pm}(v,.)$, where $K:=\Box+\xi R+m^2$ is the Klein-Gordon operator, we can use the Propagation of Singularities Theorem (Theorem 6.1.1' of \cite{Duistermaat+1972}) to propagate the null vector $(x,k)$ along the light-like geodesic that it generates to points $(y,l)$ in
$WF_{(s+1)}(E^{\pm}(v,.))\setminus WF_{(s)}(v)$. We may find such points with $y$ to the past (-) or future (+) of the support of $v$. However, since $E^{\pm}(v,.)\equiv 0$ there, this gives a contradiction. Thus we must have $WF_{(s+1)}(u)=\emptyset$.
\end{proof*}

\begin{proposition}\label{Prop_W21SA}
Let $\omega_2$ be a Hadamard two-point distribution of a free scalar field and let $T_1$ be the compression of $T$ to the one-particle Hilbert space $\mathcal{K}_{2\omega_2}$. Assume that $Q_j\equiv 1$ and $f:=\sum_{j\in J}f_j$ is real-valued. Then $T_1$ is essentially self-adjoint.
\end{proposition}
\begin{proof*}
Suppose that $\psi\in\mathcal{K}_{2\omega_2}$ is an eigenvector of $T_1^*$ with eigenvalue
$\lambda\in i\mathbb{R}\setminus\left\{0\right\}$ and consider the distribution $u(h):=\langle\psi,\kappa(h)\rangle$ on $M$. The eigenvalue equation $T_1^*\psi=\lambda\psi$ implies
$\overline{\lambda}u(h)=\langle \psi,T_1\kappa(h)\rangle=\omega_2(fu,h)$, where we used the explicit expression for $T_1$ in terms of creation and annihilation operators. Now set $v:=fu\in\mathcal{E}'(M)$, so there is an $s\in\mathbb{R}$ such that $WF_{(s)}(v)=\emptyset$. Applying Lemma \ref{Lem_WFs+} and $\overline{\lambda}v=f\omega_2(v,.)$ we find that $WF_{(s+1)}(v)\subset WF_{(s+1)}(\omega_2(v,.))=\emptyset$. Iteration gives $WF_{(s)}(v)=\emptyset$ for all $s$, i.e.\ $WF(v)=\emptyset$.

Next we define $v_1:=\mathrm{Re}(v)$ and $v_2:=\mathrm{Im}(v)$, which are real-valued test-functions. Splitting the equation $\overline{\lambda}v=f\omega_2(v,.)$ into real and imaginary parts yields
\begin{eqnarray}
-i\lambda v_2&=&f\omega_{2+}(v_1,.)-\frac{f}{2}E(v_2,.)\nonumber\\
i\lambda v_1&=&f\omega_{2+}(v_2,.)+\frac{f}{2}E(v_1,.).\nonumber
\end{eqnarray}
Inserting $v_2$ in the first line, $v_1$ in the second line and subtracting yields $0=v_1(v_1)+v_2(v_2)=\int_M |v|^2$, by the anti-symmetry of $E$ and the symmetry of $\omega_{2+}$. This means we have $v=0$ (pointwise) and hence $\overline{\lambda}u=\omega_2(v,.)=0$, which implies $\psi=0$. In other words, the range of $T_1\pm iI$ is dense for both signs, which means that $T_1$ is essentially self-adjoint.
\end{proof*}

\begin{theorem}\label{Thm_W2SA}
Let $\omega$ be a quasi-free Hadamard state of a free scalar field and $T:=\ :\pi_{\omega}(\Phi^2):(f)$ with $f\in\mathcal{S}$. Then $T$ is essentially self-adjoint on $\mathcal{D}_{\omega}$.
\end{theorem}
\begin{proof*}
The result follows from Theorem \ref{Thm_SA} combined with Proposition \ref{Prop_W21SA}. We also note that Theorem \ref{Thm_SAGL} provides an opportunity to deduce some information on the spectral projections of $\overline{T}$.
\end{proof*}

Thus we see that for any $d\ge 2$ the Wick square $:\pi_{\omega}(\Phi^2):(f)$ is essentially self-adjoint on its natural domain.\footnote{In \cite{Brunetti+} Wick polynomials are defined on a ''microlocal domain of smoothness''. This domain can be shown to be larger than the ''Wightman domain'', which we use here. A Wick polynomial may be essentially self-adjoint on the microlocal domain of smoothness without being essentially self-adjoint on the Wightman domain, a possibility which might be worthy of further investigation, also for higher Wick powers. I would like to thank Romeo Brunetti for bringing this to my attention.} To draw a similar conclusion for the components of the stress-energy-momentum tensor requires a study of its compression to the one-particle Hilbert space, which is unfortunately more complicated than for the Wick square due to the presence of derivatives.

\section{Conclusions}\label{Sec_Further}

In this paper we have shown that it is possible to obtain self-adjointness results for operators in quantum field theory by exploiting the microlocal spectrum condition. This technique allowed us to generalise previously known results to a large class of physically relevant states on all globally hyperbolic spacetimes, at for least for suitable operators. Note that we imposed no requirements on the boundary of the spacetime, such as geodesic completeness, which is usually assumed to prove the essential self-adjointness of the wave operator on the Hilbert space of $L^2$-functions, nor did we restrict the spacetime dimension $d$.

Our strategy for proving the essential self-adjointness of second order Wick polynomials was to reduce the problem to the one-particle Hilbert space and to try and exploit the essential self-adjointness of the compression $T_1$. Together with our locality lemma this opens up the way to apply estimates obtained by Verch in his study of local quasi-equivalence \cite{Verch1994}. We point out that Baez' proof \cite{Baez1989} of the self-adjointness of the Wick square in a ground state on a static spacetime also reduces the problem to the one-particle Hilbert space, but uses very different techniques. Indeed, he studies the action of commutators with the Wick square, which is determined by a linear operator $L$ on the one-particle Hilbert space. A result of Poulsen (published by Klein \cite{Klein1973}) then guarantees the essential self-adjointness of the Wick square, as soon as $L$ is a Hilbert-Schmidt infinitesimal symplectic map.

The microlocal spectrum condition and Nelson's Theorem do allow some further results, which we state here without proof, because the proofs are variations on the ones given in the main text. If the Hadamard condition is strengthened so as to exclude also time-like singularities (as is the case for free scalar fields), one may restrict the field and its derivatives to spacelike hypersurfaces. By Nelson's Theorem these restrictions are essentially self-adjoint (when suitably smeared to make them symmetric).

A different strategy to prove essential self-adjointness results would be to try and exploit the fact that the two-point distribution of a free scalar field becomes more regular in lower dimensions. Indeed, the Sobolev wave front set of a Hadamard two-pint distribution of a free scalar field in a $d$-dimensional spacetime ($d\ge 2$) is given by\footnote{This result generalises Lemma 5.2 of \cite{Junker+} and can be inferred from Theorem B.10 and Section 5.1 (especially the equation displayed above Equation (62)) loc.cit.}
\[
WF_{(s)}(\omega_2)=\left\{\begin{array}{ll}
WF(\omega_2)&\mathrm{if}\ s\ge\frac{3-d}{2}\\
\emptyset&\mathrm{if}\ s<\frac{3-d}{2}
\end{array}\right..
\]
In the extreme case $d=1$ of a scalar field in a one-dimensional ''spacetime'' (i.e.\ a harmonic oscillator) no singularities occur in the two-point distribution at all and therefore second order Wick polynomials (with derivatives) are essentially self-adjoint when they are symmetric, as may be seen from Lemma \ref{Lem_Wick} and Nelson's Theorem. In fact, a similar conclusion holds for semi-bounded fourth order Wick polynomials, by replacing Nelson's Theorem by Nussbaum's Theorem. In particular we note that the fourth Wick power $:\Phi^4:(f)$ is essentially self-adjoint when smeared with any positive measure $f$. This is in contrast to the results of \cite{Rabsztyn1989}, who found that the third Wick power in the one-dimensional setting is only essentially self-adjoint for non-generic choices of the smearing function.

For $d\ge 2$ it is harder to see if the extra regularity of the two-point distribution can lead to further self-adjointness results. For example, in $d=4$ it is known that the compression $T_1$ of a Wick square to the one-particle Hilbert space of the Minkowski vacuum in Minkowski spacetime is a bounded operator \cite{Langerholc+1965}. This leads to additional essential self-adjointness of the Wick square, also for more general smearing functions than those of class $\mathcal{S}$ and in a large class of non-quasi free states, using Nelson's Theorem (as in Proposition \ref{Prop_NelsonFock}). We believe the boundedness of $T_1$ for a Wick square can be generalised to curved spacetimes for $d=2$, but for $d\ge 3$ the situation is less clear. In the presence of derivatives one would generally expect to require smaller $d$ in order to obtain similar results. (Note on the other hand that Lemma \ref{Lem_WFs+} is independent of $d$.)

${}$\\[12pt]
{\bf Acknowledgements}\\[6pt]
I would like to thank Robert Wald for encouraging remarks and Chris Fewster for posing a critical question during a
special programme ''QFT on curved spacetimes and curved target spaces'', held at ESI, Vienna in Spring 2010. A first
draft of this paper was prepared at the Institute for Theoretical Physics at the University of G\"ottingen and was
supported by the German Research Foundation (Deutsche Forschungsgemeinschaft (DFG)) through the Institutional Strategy
of the University of G\"ottingen and the Graduiertenkolleg 1493 ''Mathematische Strukturen in der modernen
Quantenphysik''.

\end{document}